\documentclass{rstransa}
\jname{rsta}
\Journal{Phil. Trans. R. Soc}

\usepackage{amsmath,amssymb,amsfonts,amsthm}
\usepackage{graphicx}
\usepackage{times}
\usepackage{helvet}
\usepackage{courier}
\usepackage{bm}
\usepackage{dcolumn}
\usepackage[none]{hyphenat}
\usepackage{url}
\usepackage{caption}
\usepackage{subcaption}
\usepackage[toc,page]{appendix}

\usepackage{setspace}
\usepackage{float}

\usepackage[utf8]{inputenc} 

\graphicspath{ {plots/},{style_plots/} }

\usepackage{color}

\renewcommand \thesection{\arabic{section}}

\newcommand{\fScl}{\ensuremath{\Lambda}}
\newcommand{\tScl}{\ensuremath{\mu}}
\newcommand{\xScl}{\ensuremath{\nu}}

\newcommand{\mBH}{\ensuremath{M_{\rm BH}}}
\newcommand{\rTort}{\ensuremath{r_*}}
\newcommand{\mpl}{\ensuremath{m_{\rm pl}}}
\newcommand{\mT}{\ensuremath{m_{\rm T}}}
\newcommand{\rs}{r_*}
\newcommand{\msun}{\ensuremath{M_\odot}}
\newcommand{\p}{\partial}

\newcommand{\figref}[1]{Fig.~\ref{#1}}

\begin{document}

\title{Accretion of a Symmetry Breaking Scalar Field by a Schwarzschild Black Hole}

\author{
Dina Traykova$^{1}$, Jonathan Braden$^{1}$ and Hiranya V. Peiris$^{1,2}$}

\address{$^{1}$Department of Physics and Astronomy, University College London, London WC1E 6BT, U.K.\\
$^{2}$The Oskar Klein Centre for Cosmoparticle Physics, Stockholm University, AlbaNova, Stockholm, SE-106 91, Sweden}
\subject{Astrophysics, Cosmology, High Energy Physics, Relativity}


\corres{Dina Traykova\\
\email{dina.traykova.13@alumni.ucl.ac.uk}}

\begin{abstract}
  We simulate the behaviour of a Higgs-like field in the vicinity of a Schwarzschild black hole using a highly accurate numerical framework.
  We consider both the limit of the zero-temperature Higgs potential, and a toy model for the time-dependent evolution of the potential when immersed in a slowly cooling radiation bath.
  Through these numerical investigations, we aim to improve our understanding of the non-equilibrium dynamics of a symmetry breaking field (such as the Higgs) in the vicinity of a compact object such as a black hole.
  Understanding this dynamics may suggest new approaches for studying properties of scalar fields using black holes as a laboratory.
\end{abstract}


\begin{fmtext}

\end{fmtext}


\maketitle

\section{Introduction}
Scalar fields are ubiquitous ingredients in theories of fundamental physics beyond the Standard Model.
For example, the Higgs field is responsible for electroweak symmetry breaking, resulting in the generation of masses of quarks, leptons, and W$\pm$ and Z$^0$ gauge bosons.
The QCD axion field provides a potential solution to the strong CP problem of quantum chromodynamics~\cite{PhysRevLett.38.1440},
while simultaneously providing a cosmological dark matter candidate~\cite{PRESKILL1983127}.
Another important hypothetical field in cosmology is the inflaton, which is thought to have driven the accelerated expansion of the early universe~\cite{Guth:1980zm,Linde:1983gd,Starobinsky:1980te} and produced the density fluctuations that seeded all of the observed structure in the Universe~\cite{Mukhanov:1981xt,Mukhanov:1990me}. 
Additionally, the current accelerated expansion of the Universe may be the result of the vacuum energy of yet another scalar field, a scenario known as quintessence~\cite{Caldwell:1997ii}.
Finally, compactifications of higher dimensional ambient spacetimes of string theory to the $3+1$ dimensional spacetime consistent with our Universe lead to a wide variety of moduli and axionic scalar fields~\cite{Kachru:2003aw}.

However, despite their prevalence as theoretical tools, among these only the Higgs has been observed experimentally~\cite{Chatrchyan:2012xdj,Aad:2012tfa}.
Assuming a given scalar exists, the lack of experimental verification may be a result of either extremely weak interactions with the particles of the Standard Model or the high mass scale of the particle excitations of the field.
However, even if their interactions with the known particles are extremely weak and thus unamenable to direct detection by terrestrial experiments, the equivalence principle implies that they must interact gravitationally. 
Hence, even in the absence of electromagnetic, strong or weak interactions, we might be able to constrain some of their properties through their interaction with objects with extreme gravitational fields, such as black holes. 
This may even be the only environment in which some scalars could be studied.
However, in order to realise this possibility, we must be able to accurately model the dynamical evolution of these fields in extreme gravitational environments.
Recently, it has been suggested that axionic fields permeating the whole of spacetime may be detected using observations from gravitational wave detectors such as advanced LIGO through the effects of black hole superradiance (e.g.,\,\cite{Arvanitaki:2014wva,Arvanitaki:2016qwi}). 
In this work, we take an additional step towards understanding scalar field interactions with black holes by modelling the evolution of a Higgs-like field near a Schwarzschild black hole.

Previous investigations of the gravitational interactions between black holes and scalar fields have primarily focussed on questions of black hole formation and emergent critical phenomena during the collapse of local scalar field overdensities~\cite{Choptuik:1992jv,Gundlach:2007gc,Clough:2016jmh}.
Additional studies have also looked at the emergence of an inflationary expansion from highly inhomogeneous initial conditions (e.g.,\,~\cite{Goldwirth:1989pr,KurkiSuonio:1993fg,East:2015ggf,Clough:2016ymm}).
These investigations require the full machinery of numerical relativity, where spacetime must be treated as a dynamical entity strongly coupled to the evolving scalar field~\cite{Baumgarte:2010:NRS:2019374}.
A seminal achievement in this field was the simulation of binary black hole mergers~\cite{Pretorius:2005gq}.

The dynamics of scalar field accretion onto black holes has also been studied for many years. 
Much research has been focused on the late-time power law behaviour of scalar fields evolving around black holes.
First discovered by Price~\cite{PhysRevD.5.2419}, who modelled a scalar field in the exterior of a Schwarzschild black hole, this behaviour has been studied actively, mainly in the context of a Kerr black hole~\cite{Barack:1999ma, Hod:1999ci, Burko:2013bra}.
The majority of the research has been focused on the exterior black hole spacetime; however, in more recent years there has been increasing interest in the interior spacetime \cite{PhysRevLett.63.1663, PhysRevD.41.1796, Ori:1991zz, Brihaye:2016vkv}.

In a recent study \cite{Thuestad:2017ngu}, the late-time behaviour of scalar fields has been investigated in both the exterior and the interior spacetime of a rotating Kerr and a static Schwarzschild black hole.
Since this study focuses on both the exterior and interior regions, the authors have chosen to use a Kerr spacetime generalisation of the Eddington-Finkelstein coordinates, which are smooth at the horizon of a Schwarzschild black hole.
The Teukolsky equation is solved using numerical techniques as described in Ref.\ \cite{Zenginoglu:2011zz}.
They find that in the interior of a Kerr or a Schwarzschild black hole a detector would record a finite number of oscillatory cycles in the field before it reaches the black hole singularity. 

The interaction of scalar fields with rotating black holes has also been used to study the instabilities in such systems.  
For example, in Ref.\ \cite{Witek:2012tr} the authors model the evolution of linearised fields in the vicinity of a rotating black hole, and they demonstrate the existence of highly non-trivial time dependence and even rapid growth.
In another study the possible mechanisms which can destabilise a Kerr black hole in the presence of a scalar field have been investigated \cite{Cardoso:2013opa}.
It was shown that matter surrounding a black hole could cause two types of instabilities depending on the sign of the effective mass squared of the scalar.
For a negative square of the effective mass, the scalar field around the black hole becomes spontaneously excited and the spacetime develops non-trivial scalar hair.
Positive effective mass squared, on the other hand, is shown to lead to a superradiant instability, which affects Kerr black hole solutions.

There exist multiple proofs that static Schwarzschild black holes do not support regular scalar hair \cite{PhysRevD.5.1239, Faraoni:2013iea}. 
However, it has been proven possible to have scalar hair outside the event horizon in the form of non-trivial scalar profiles.
For instance, two types of quantum hair are described in Ref.\ \cite{Coleman:1991ku}, one associated with the discrete gauge charge and another with the colour magnetic charge.
In another study it was shown that black holes can support non-trivial hair in the form of Abelian Higgs vortices \cite{Achucarro:1995nu}.
In theories of massive gravity, black holes have also been shown to support massive graviton hair \cite{Brito:2013xaa}.

The Schwarzschild solution is particularly useful when we are only interested in the final state of evolution of the black hole.
In a study mentioned previously~\cite{Cardoso:2013opa}, the authors use the significantly simpler Schwarzschild background solution to study the final state of spontaneous scalarisation.
In Ref.\ \cite{Barranco:2012qs} the evolution of scalar fields coupled to a Schwarzschild black hole has been modelled.
It was shown that non-trivial scalar field profiles can survive in the vicinity of black holes for cosmological times for a certain range of black hole and scalar field masses. 
The equations of motion are evolved numerically using second-order derivative operators satisfying summation by parts and a third-order Runge-Kutta operator for the time integration, as described in Ref.\ \cite{Megevand:2007uy}.
Finally, in Ref.\ \cite{Frolov:2017asg}, the accretion of a scalar field responsible for screening mechanisms by a static Schwarzschild black hole has been modelled using spectral methods and Gauss-Legendre time integration.  It is shown that for certain parameter choices, the equilibrium accretion profile is dynamically unstable, and can even lead to the creation of a naked singularity.

Much of this past work has focussed on static or nearly static situations, while we will explore a strongly nonequilbrium dynamical time evolution of the scalar field.  We will work in the test field limit, where the backreaction of the scalar field on the black hole spacetime is ignored.
However, we fully account for the evolution of the scalar field within this externally provided spacetime.
Furthermore, we restrict ourselves to the spherically symmetric Schwarzschild black hole.
This should provide a reasonable approximation as long as the integrated flux of energy falling through the black hole horizon is much less than its initial mass,
and provides a much simpler arena in which to carry out the exploratory investigation performed here. 

We developed a numerical code to approach this problem that makes use of Chebyshev-based pseudo-spectral methods~\cite{boyd01,Grandclement:2007sb} for performing numerical differentiation, and a tenth order accurate Gauss-Legendre method~\cite{Butcher1964,Braden:2014cra} for the time evolution.
Our approach is able to produce a highly accurate evolution for a wide variety of scalar fields within a spherically symmetric black hole spacetime.
Here, we will apply it to the question of dynamical symmetry breaking by a Higgs-like field to study the impact of the gravitational forces on the Higgs' ability to reach its true vacuum minimum.
For all of the field evolutions presented in this paper, we performed convergence testing in both the time-step and pseudospectral order and verified the results were accurate to $\mathcal{O}(10^{-8})$.  Although we have not implemented them here, the use of a pseudospectral discretisation also makes it easy to implement perfectly-matched layer boundary conditions to absorb outgoing radiation from the simulation, which can again be easily tuned to achieve machine precision.
Since the results in this paper are intended as a proof-of-concept, the ability to easily extend the code to cases with emitted radiation is an important consideration.

Before we consider a time-dependent model for the symmetry breaking dynamics, where the potential evolves from one with a minimum to a maximum at the origin during the course of the simulations, we first consider the two limiting cases for this transition.
We start with the case of the static zero temperature double-well potential, with the field initially starting near the local potential maximum at the origin.
This provides a simple model for the dynamics following a quench of the potential.
We then consider the case where the field is instead stabilised to have a minimum around the origin.
We find qualitatively similar behaviour to the original symmetry breaking case, with the field undergoing oscillations that pile up near the horizon.
Finally we present the results for the time-dependent model.
Initially, the field oscillates around the origin, but as the potential transitions from the symmetry restoring to the symmetry breaking phase, qualitatively new behaviour emerges as a series of spherical domain wall shells form around the singularity, with the field undergoing oscillations in between the walls.
This intriguing phenomenon develops from a mismatch between the constant field and constant potential shape surfaces, which we have imposed by hand in this work.
We leave exploration of a fully dynamical mechanism to induce this mismatch to future work.

\section{Scalar field equation of motion}
Although black holes are dynamically evolving objects, static black hole solutions are a valid approximation when studying the accretion of a scalar field.
It has been shown in that it takes approximately $10^7$ years to change the mass of an accreting supermassive black hole by 1 $\%$~\cite{Frolov:2017asg}. The time scale relevant to the evolution of the Higgs field is 
\begin{equation}
\Delta t \sim m_{\rm H}^{-1}\,,
\end{equation}
where $m_{\rm H}$ is the mass of the Higgs. 
This is significantly smaller than the time necessary for the black hole mass to change, therefore we may treat the black hole itself as non-dynamical.
We will make use of this approximation throughout.
Further, we will ignore the effects of the cosmological expansion, and treat the asymptotic behaviour of our spacetime as Minkowski.
The assumption of spherical symmetry and Minkowski asymptotic behaviour leads to the well-known Schwarzschild solution
\begin{equation}
    ds^2 = -\left(1-\frac{2\alpha}{r}\right)dt^2 + \left(1-\frac{2\alpha}{r}\right)^{-1}dr^2 + r^2d\Omega^2,
    \label{equ:met}
\end{equation}
where $\alpha=G\mBH$, $\mBH$ is the mass of the black hole, and we have chosen units with the speed of light $c=1$.
The radial coordinate $r$ is known as the areal coordinate since a sphere of radius $r$ has area $4\pi r^2$.
Unfortunately, this coordinate system is ill-suited to numerical evolution since as $r$ approaches the Schwarzschild radius at $r_{\rm S} = 2\alpha$, the ingoing waves of the field will become extremely blueshifted.
The high spatial frequency of the waves near the horizon then leads to extreme numerical resolution issues when attempting to evolve the field on a discrete lattice.
To avoid this problem, it is therefore convenient to transform to tortoise coorinates $r_*$ defined as
\begin{equation}
  \rs = r + 2\alpha\ln\left(\left|\frac{r}{2\alpha}-1\right|\right) \qquad \mathrm{with} \qquad
  \frac{d\rs}{dr}=\left(1-\frac{2\alpha}{r}\right)^{-1} \,.
  \label{equ:tor}
\end{equation}
These coordinates map the horizon at $r_{\rm S} = 2\alpha$ to $-\infty$ at an appropriate rate so as to cancel the undesirable blueshifting effect in wavelength of ingoing waves. 

In the new coordinates the metric is
\begin{equation}
    ds^2=-\left(1-\frac{2\alpha}{r}\right)dt^2+\left(1-\frac{2\alpha}{r}\right)d{\rs}^2+r^2d\Omega^2\,.
    \label{equ:met-tor}
\end{equation}
The corresponding equation of motion is derived in Appendix~\ref{app:EOM} and has the form
\begin{equation}
\frac{\p^2\phi}{\p t^2} = \frac{\p^2\phi}{\p{\rs}^2} + \frac{2}{r}\left(1-\frac{2\alpha}{r}\right) \frac{\p\phi}{\p\rs} - \left(1-\frac{2\alpha}{r}\right)\frac{dV}{d\phi} \,,
\label{equ:eom_tor}
\end{equation}
where $\phi$ is the field amplitude and $V$ is its potential.
Details of our numerical approach to solving this equation may be found in Appendix~\ref{app:num}.

\section{Accretion of a scalar with a symmetry breaking potential}
We now study the accretion of a Higgs-like field as it undergoes a symmetry breaking transition and rolls from a location near the origin to its final vev at the symmetry breaking minimum of the potential.
In particular, we have in mind a situation where the field is immersed in the hot thermal plasma of the early universe, with temperature $T$.
As the universe expands and cools, the finite temperature potential evolves from one with a minimum at the origin, which traps the field and prevents symmetry breaking, to a maximum at the origin, which allows the field to roll away from the origin initiating the process of symmetry breaking.

If the typical size of the fluctuations (quantum or thermal) in the field is much larger than its mean displacement from the maximum, then a whole set of long-wavelength modes will experience a tachyonic instability.
  These fluctuations will not respect spherical symmetry, and therefore a full $3+1$ dimensional approach must be adopted to study that situation.
  To avoid this complication, we will consider situations in which the field is initially displaced from the origin.
  
Rather than study the somewhat complicated dynamics for the full finite-temperature effective potential of the Standard Model Higgs, here we consider a toy potential of the form
\begin{equation}
  V(\phi) = \frac{\lambda}{4}\left(\phi^2-\phi_0^2\right)^2 + \frac{\mT^2}{2}\left(\frac{T}{T_0}\right)^2\phi^2 \, ,
  \label{equ:pot_td}
\end{equation}
which has explicit time dependence through the temperature $T$.
Another departure from the Standard Model Higgs is that this potential supports domain walls.
For numerical simulations, it is convenient to work in rescaled coordinates where $\alpha=1$ and to measure the field in units of $\phi_0$. 
After some manipulations (see Appendix~\ref{app:num}) we find that our dimensionless time and space coordinates are $\bar{t} = \mu t$ and $\bar{r}^* = \mu r^*$, with $\mu = \mpl^2/\mBH$.
The corresponding dimensionless potential is
\begin{equation}
  \bar{V}(\bar{\phi}) = \frac{\bar{\lambda}}{4}\left(\bar{\phi}^2-1\right)^2 + \frac{\bar{m}_{\rm T}^2}{2}\left(\frac{T}{T_0}\right)^2\bar{\phi}^2 \, ,
\end{equation}
where the overbar denotes the dimensionless equivalents of the variables in~\eqref{equ:pot_td}.
Explicitly
\begin{equation}
  \bar{\lambda} = \lambda \left(\frac{\phi_0}{\mu}\right)^2 = \lambda\left(\frac{\phi_0}{\mpl}\right)^2\left(\frac{\mBH}{\mpl}\right)^2 
\end{equation}
and
\begin{align}
    \bar{m}_{\rm T}^2 & = \left(\frac{m_T}{\mu}\right)^2= \sigma\lambda\left(\frac{\phi_0}{\mu}\right)^2 = \sigma\,\bar{\lambda} \, ,
\end{align}
where $\sigma$ adjusts the shape of the potential and controls the transition from the symmetry restoring to symmetry breaking phase. 
For $\sigma(T/T_0)^2 > 1$ the origin $\phi=0$ is a local minimum of the potential, while for $\sigma(T/T_0)^2 < 1$ it is a local maximum.
We can interpret these new potential terms as finite temperature corrections, and since we are interested in the regime where the Higgs is initially stabilised at the origin, in this section we focus on the regime $m_T^2 \gtrsim\lambda\phi_0^2$.
For definiteness, we take $\sigma=2.5$ here.

Before proceeding to numerical simulations, we must provide a connection between the temperature parameter $T$ and the time-variable $\bar{t}$ appearing in our code.
In order to provide this connection, we adopt the simplistic assumption that the Schwarzschild time $t$ used in our simulations is equal to the cosmic time $t_{\rm cos}$ used to describe the plasma in the early universe.
In reality, we do not expect this equivalence to hold due to, for example, time dilation effects near the horizon of the black hole.
A proper accounting would require taking the asymptotic behaviour of our metric to be that of the ambient Friedmann-Robertson-Walker (FRW) spacetime, as well as dynamically evolving the thermal plasma throughout the spacetime.
However, in the interest of simplicity we adopt the approximation above.
In a more careful treatment, we might expect the temperature to drop more slowly near the horizon, which will further act to increase the time scale over which the field experiences symmetry breaking in the near-horizon regime.
In Section~\ref{sec:t-vary-pot} we will explore one consequence of imposing this relationship between temperature and time.
 
For relativistic particles the temperature is inversely proportional to the scale factor.
Assuming the universe is radiation dominated, the relation between the scale factor and time is given by the Friedmann equation
\begin{equation}\label{eqn:friedman}
    H^2\equiv\left(\frac{\dot{a}}{a}\right)^2=\frac{8\pi}{3\,\mpl^2}\rho_{\rm tot}\,,
\end{equation}
where $\rho_{\rm tot}$ is the energy density of the universe.
For relativistic particles, we have
\begin{equation}\label{eqn:rad-density}
  \rho_{\rm tot}=\frac{\pi^2}{30}g_{\rm eff}T^4
\end{equation}
where
\begin{equation}
    g_{\rm eff}=\sum\limits_{i={\rm bosons}}g_i+\frac{7}{8}\sum\limits_{j={\rm fermions}}g_j\, ,
\end{equation}
and $g_i$ and $g_j$ are the number of relativistic (i.e., $m \ll T$) bosonic and fermionic degrees of freedom, respectively. 
At $T\sim 1\,\mathrm{TeV}$ we can take all the particles of the Standard Model as being in the relativistic regime to obtain
\begin{equation}
    g_{\rm eff}(T\sim 1\,\mathrm{TeV}) = 28+\frac{7}{8}90 = 106.75\,.
\end{equation}
During our simulations, the temperature decays by roughly a factor of two, so for the purposes of this work we will treat $g_{\rm eff}$ as a constant in time.
Substituting the expression for the energy density~\eqref{eqn:rad-density} into the Friedmann equation~\eqref{eqn:friedman}, we find
\begin{equation}
\frac{T}{T_0}=\left[2\left(\frac{T_0}{\mpl}\right)^2\frac{\mBH}{\mpl}\left(\frac{8 \pi^3}{90}g_{\rm eff}\right)^{1/2}\left(\bar{t}-\bar{t}_0\right)+1\right]^{-1/2}\, .
\end{equation}

\begin{figure}[H]
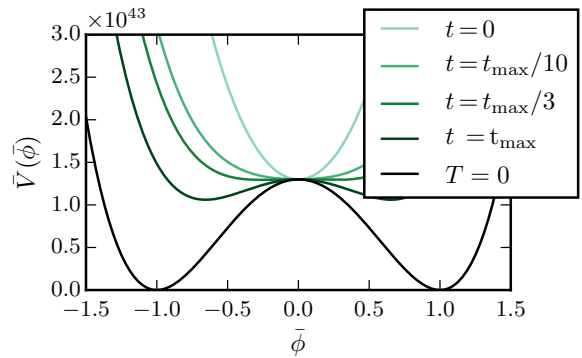

  \centering
  \includegraphics[width=0.6\textwidth]{{{pot_td2}}}
  \caption{Evolution of the scalar field potential, as given in~\eqref{equ:pot_td} with the temperature $T$ given by~\eqref{equ:td}.  We have defined $t_{\rm max}=2\,{\rm GeV}^{-1}$, which corresponds to the duration of the numerical simulations we present below. The transition of the potential from one with a minimum to one with a maximum at the origin occurs at $t\sim t_{\rm max}/2$ and $T\approx0.63\,T_0$. For reference, we also show the case where $T=0$.}
  \label{fig:potential-evolution}
\end{figure}

Motivated by this, we will consider a temperature-time relationship of the form
\begin{equation}
  \frac{T}{T_0} = \left(\beta\left(\bar{t}-\bar{t}_0\right)+1\right)^{-1/2} \, ,
  \label{equ:td}
\end{equation}
with the adjustable parameters $\beta$ and $t_0$ chosen so that the potential undergoes a symmetry breaking transition during the course of the simulation.
This breaks the more precise relationship with the cosmological evolution of our own Universe,
but will still serve to model the essential dynamics of the symmetry breaking process. 
The temperature $T_0$, taken here to be around 1 TeV is a reference value, used to track the temperature change from a phase where the scalar field is trapped at the origin ($T \gg T_0$) to one where it undergoes spontaneous symmetry breaking ($T \ll T_0$).
Since we are primarily interested in this transitionary regime, we start the simulations with $T \sim T_0$ but such that the potential has a minimum at the origin.
The initial potential along with its subsequent evolution is illustrated in~\figref{fig:potential-evolution}.
We wish to study the dynamical transition from a symmetry restored to symmetry broken state for the field, so we first consider the two regimes separately.

\subsection{Zero-temperature case}\label{sec:zero-temp-pot}
\begin{figure}[H]
    \centering
    \begin{subfigure}[t]{0.48\textwidth}
        \includegraphics[width=\textwidth]{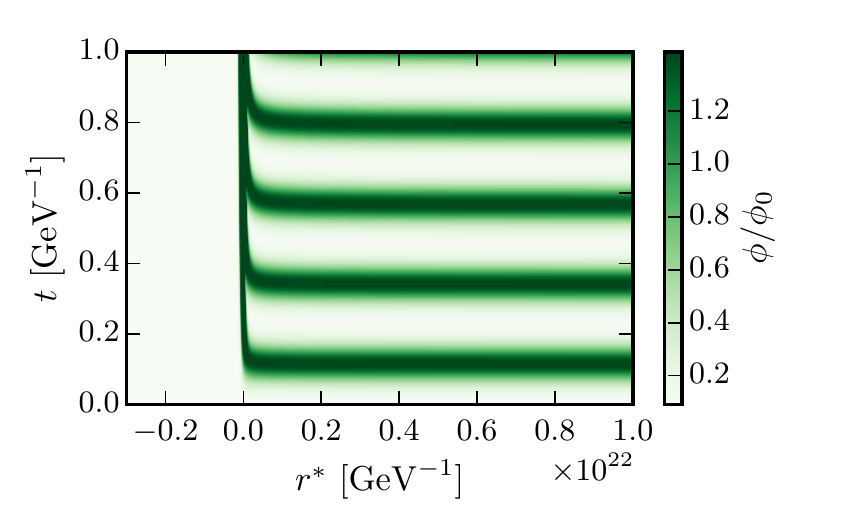}
    \subcaption{BH mass, $\mBH=1\,\msun$.}
    \label{fig:m1}
    \end{subfigure}
    \,
    \begin{subfigure}[t]{0.48\textwidth}
        \includegraphics[width=\textwidth]{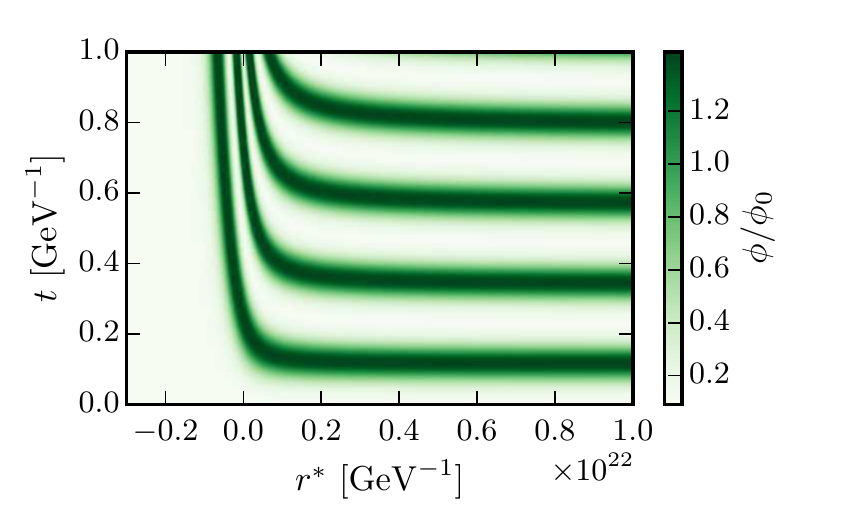}
        \subcaption{BH mass, $\mBH=10\,\msun$.}
        \label{fig:m10}
    \end{subfigure}
    \\
    \centering
    \begin{subfigure}[t]{0.48\textwidth}
        \includegraphics[width=\textwidth]{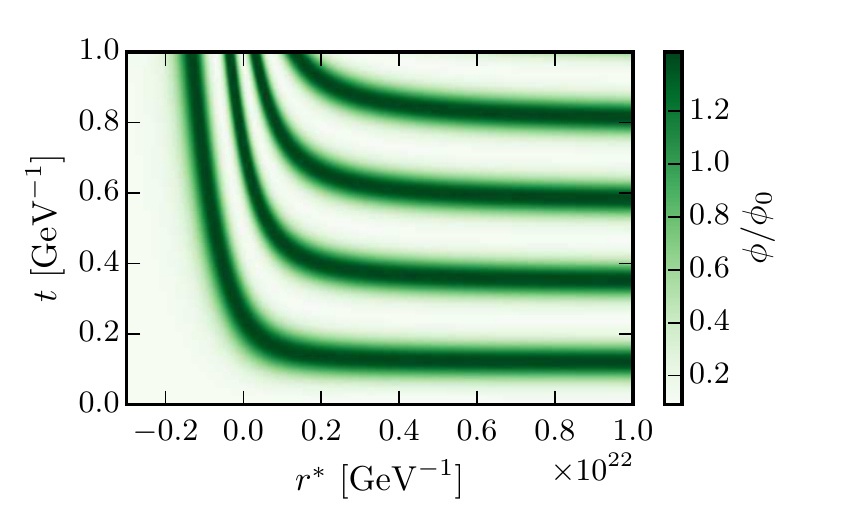}
        \subcaption{BH mass, $\mBH=20\,\msun$.}
        \label{fig:m20}
    \end{subfigure}
    \,
    \begin{subfigure}[t]{0.48\textwidth}
        \includegraphics[width=\textwidth]{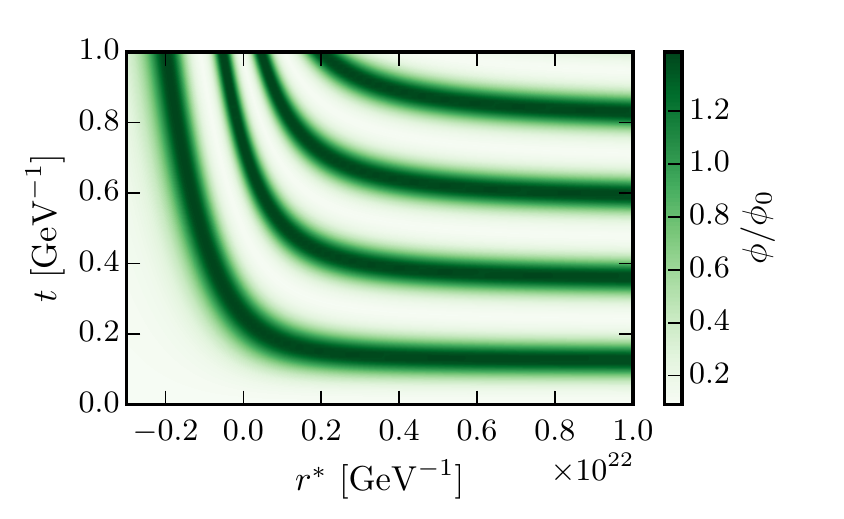}
        \subcaption{BH mass, $\mBH=30\,\msun$.}
        \label{fig:m30}
    \end{subfigure}
    \caption{Evolution of the field in the potential~\eqref{equ:pot_td} with zero temperature, illustrated by the line $T=0$ in~\figref{fig:potential-evolution}, and homogeneous initial condition $\phi_i=0.1\phi_0$ for a range of black hole masses.  In each case, $r^* = 0$ corresponds to the an areal coordinate at a fixed fraction of the Schwarzschild radius $\frac{r}{2\alpha} \sim 1.279$.}
    \label{fig:ti}
\end{figure}

We first present the dynamics of the field following an instantaneous quench of the potential.
We model this situation by setting the finite temperature correction terms of the potential to zero.
This case is presented in~\figref{fig:potential-evolution} labelled as $T=0$ .
We now study the symmetry breaking dynamics of the field, assuming it starts from an initially homogeneous configuration of amplitude $\phi_i$.
Since we are interested in the symmetry breaking dynamics, this should be much smaller than $\phi_0$.
However, decreasing the initial field value leads to an increasing time delay before the field rolls off the hill.
To avoid this initial transient and a spinodal instability of the quantum vacuum, we focus on the field oscillations around the minimum and choose $\phi_i/\phi_0=0.1$.
In~\figref{fig:ti} we present the resulting field evolution for black hole masses of $1, 10, 20$, and $30$ $\msun$.
In order to observe the dynamics far from the black hole, we must resolve several field oscillations in the large-$r_*$ regime.
To achieve this, we evolve the system from $t\sim 0\,{\rm GeV}^{-1}$ to $t = 2\,{\rm GeV}^{-1}$. The plots below show the evolution up to $t = 1\,{\rm GeV}^{-1}$.
As expected, we can see that in all four cases the field evolution is equivalent far away from the event horizon where the spacetime approaches Minkowski space.
Moving inward towards the horizon, the gravitational effects lead to significant deviations from the $\rs \to \infty$ behaviour in our choice of coordinates.
In particular, in the simulation time coordinate, the field lingers at its original location and does not undergo symmetry breaking near the horizon for the duration of the simulation.  This is of course due (at least in part) to our choice of simulation time coordinate and the large time-dilation effects (including the spatial blueshifting) that occurs as we enter the near horizon regime.
For different black hole masses this transition should happen at different coordinate distances, given that this is determined by the $2G\mBH/r$ term in the metric (see~\eqref{equ:met}) being much less than one.
If we assume that this effect will become important when $2G\mBH/r\lesssim0.1$, then we can estimate this transition to occur at around $0.02\times10^{22}\,{\rm GeV}^{-1}$ when $\mBH=1\,\msun$ and at $0.2\,,0.4\, \text{and}\,0.6\times10^{22}\,{\rm GeV}^{-1}$ for the $10$, $20$ and $30$ $\msun$ black holes respectively.
From the plots above we can also see that the effects of the black hole on the scalar field evolution occur more gradually as a function of our coordinate $\rs$ as we increase the mass of the black hole.

\subsection{High temperature case}\label{sec:high-t-pot}
Next we consider the other constant temperature regime we are interested in, with $T$ fixed at $T_0$ throughout the simulation so that the field is trapped near the origin.
Here we take $T_0 = 1 $TeV to be above the symmetry restoration temperature, so that the shape of the potential is given by the line labelled $t=0$ in~\figref{fig:potential-evolution}.

\begin{figure}[H]
    \centering
        \includegraphics[width=0.75\textwidth]{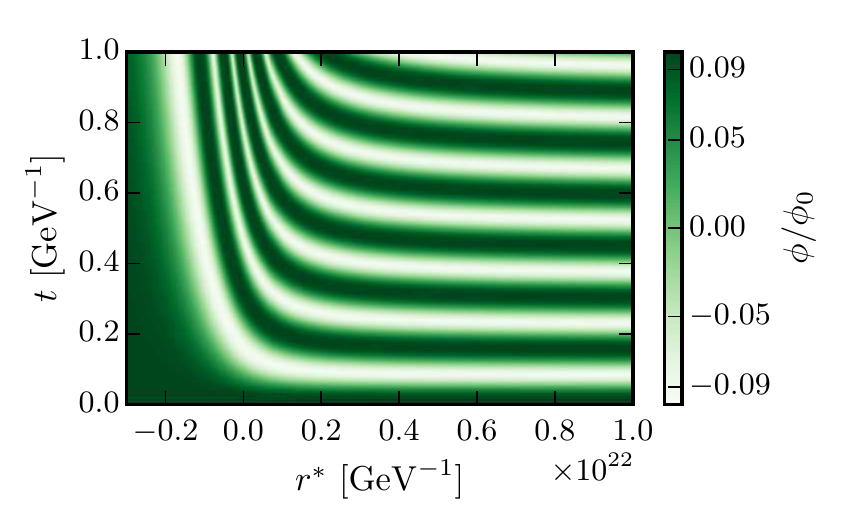}
     \caption{Evolution of the scalar field in the potential~\eqref{equ:pot_td} with finite constant temperature, corresponding to the $t=0$ line in ~\figref{fig:potential-evolution}, potential parameters $\sigma=2.5$, black hole mass $\mBH = 20\,\msun$, and homogeneous initial condition $\phi = 0.1\phi_0$.}
     \label{fig:rt_ti_comp}
\end{figure}

The resulting evolution is illustrated in~\figref{fig:rt_ti_comp} for the choice $\sigma = 2.5$, $\mBH = 20\msun$ and the same initial conditions as in Section~\ref{sec:zero-temp-pot}.
Now the field, initially at $\bar{\phi}=0.1$, rolls towards $\bar{\phi}=0$ and then continues to oscillate in an approximately quadratic potential.
For our parameter choices, the curvature of the potential around the origin is greater than around the zero temperature minima, so the field oscillates with a somewhat higher frequency.  Further, the initial displacement away from the minimum is smaller, so that the field oscillates with a smaller amplitude than before, but otherwise exhibits the same qualitative behaviour.

\subsection{Time-varying temperature case}\label{sec:t-vary-pot}
Having considered the static zero temperature case in Section~\ref{sec:zero-temp-pot}, and a static high temperature case in Section~\ref{sec:high-t-pot},
we now consider a scenario in which the potential evolves dynamically in time between these two regimes.

\begin{figure}[H]
    \centering
        \includegraphics[width=0.75\textwidth]{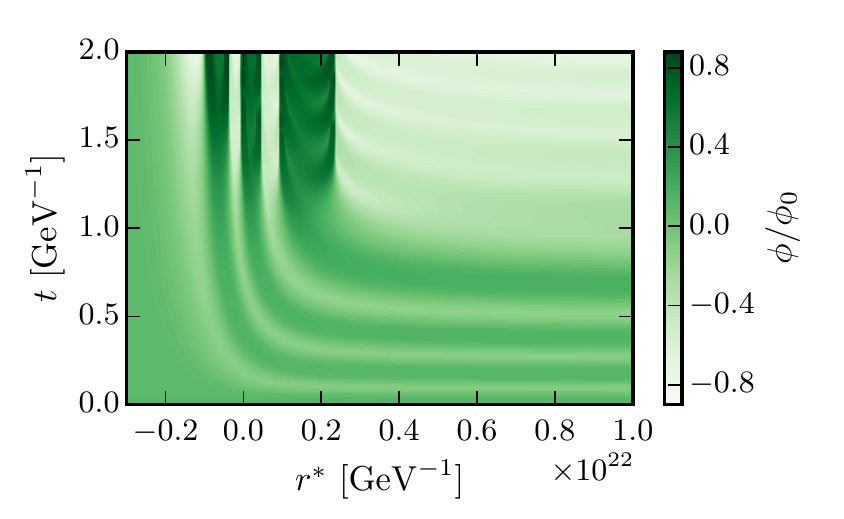}
    \caption{Field evolution in the potential~\eqref{equ:pot_td} with a time dependent temperature~\eqref{equ:td}, illustrated in~\figref{fig:potential-evolution}. As before the potential parameters are set to $\sigma=2.5$, and the black hole mass is $\mBH=20\,\msun$.}
    \label{fig:rt_td_5_0}
\end{figure}
In~\figref{fig:rt_td_5_0} we show the evolution of the field as the temperature of the universe decreases from $T\sim T_0$ to $T\sim T_0/2$. 
When comparing to previous plots, note that the scale of the $y$-axis is different to accomodate the transition of the potential which occurs at around $t=1\,{\rm GeV}^{-1}$.
At the start of the evolution the field begins to oscillate in the nearly quadratic high temperature potential with a slowly increasing amplitude and decreasing frequency.
This can be compared to~\figref{fig:rt_ti_comp}, which shows the evolution in the case of constant temperature $T=T_0$ and has the same shape as the initial time slice of the evolution in~\figref{fig:rt_td_5_0}.
As time progresses and the temperature decreases, the thermal mass correction decreases and the potential goes from one with a minimum at $\bar{\phi}=0$ to one with a maximum at the origin as illustrated in~\figref{fig:potential-evolution}.
Initially, as the potential becomes shallower, the field is able to oscillate to larger field values at lower energy cost, leading to a gradual increase in the amplitude and period of the oscillations.
When the transition of the origin from a local minimum to a local maximum occurs, rather than oscillate the field begins to roll towards one of the new minima.
Due to the nontrivial vacuum structure of the field and the presence of multiple vacua in the potential, after the symmetry breaking the field forms a series of spherical domain walls encircling the black hole.

This dynamics is significantly more complicated than for the Standard Model Higgs, but represents a novel property of potentials with discrete vacua.
It is clear that the formation of the domain walls occurs because of a mismatch in the spatial slices on which the oscillations of the $\phi$ field are homogeneous and the spatial slices on which the potential has a fixed shape (here parameterised by the temperature).
  In a more complete investigation, the order parameter controlling the shape of the potential (i.e.\ the temperature) can be evolved dynamically instead of being directly tied to the simulation time coordinate.
  Two obvious cases are to dynamically evolve the plasma temperature using relativistic hydrodynamics coupled to the black hole, or to introduce a second scalar field to act as a dynamical parameter.
We leave exploration of this interesting question to future work.
Since the introduction of an additional dynamical field allows for the presence of isocurvature as well as adiabatic modes, we expect that dynamical situations exist where the splitting of the constant $\phi$ and constant potential shape surfaces occurs.

\section{Discussion}
We developed a numerical framework to evolve relativistic scalar fields in an external black hole spacetime, with the goal of studying accretion of the field by the black hole.
We then applied this to the symmetry breaking dynamics of a Higgs-like field with a homogeneous initial value near the symmetry restoring origin.
First we considered the dynamics in the potential with temperature dependent terms set to zero and parameters given by that of the Standard Model Higgs, which corresponds to a static symmetry breaking potential.
Starting with a homogeneous initial field configuration displaced from the origin, we found that the field underwent oscillations about the vacuum state far from the black hole.
Near the black hole horizon, we instead found that in our simulation coordinates the strong effects of time dilation and gravitational interactions led to an extreme slowdown of the field oscillations and a transformation of the spatially uniform oscillations far from the black hole 
into ingoing waves as they approached the black hole horizon.
Next, we stabilised the field at the origin, rather than at a nonzero symmetry breaking value, by setting the temperature in the potential to a finite constant value.
We observed similar behaviour to the zero temperature case, except with the oscillations now occuring around the origin.
Finally, we considered the evolution during a time-dependent transition between the two regimes, and found a qualitatively new feature with the emergence of a collection of spherical domain wall shells forming around the black hole.

Overall, the results presented here serve as a proof of concept for our numerical tools to study scalar fields interacting with a black hole in the test field limit.
For all of the results presented, we verified that the maximal pointwise change in the field values from either an increase in the order of the pseudospectral interpolation, or a decrease in the time step was less than $\mathcal{O}(10^{-8})$, with the expected $dt^{10}$ and exponential convergence as the timestep and spatial interpolation order were changed.
Furthermore, the computational cost to achieve $\mathcal{O}(10^{-15})$ accuracy is minimal (at most a factor of a few), due to the high order of both the spatial and time discretisations, and if desired the use of quad precision arithmetic allows for even more accurate calculations.
Already in this simple setup, we demonstrated the possibility for strong departures from the evolution that would occur in Minkowski space.
In particular, the formation of domain walls from a homogeneous initial state is a novel departure from the behaviour expected from considerations in Minkowski space.
Unfortunately, the simple investigations performed here, with the temperature directly coupled to the simulation time coordinate, do not allow us to determine if a physical mechanism exists that can lead to the constant $\phi$ and constant potential shape surfaces (which we parameterised by $T$) being mismatched from each other.
A proper dynamical investigation including a physically evolving order parameter (such as a locally determined plasma temperature or second scalar field) requires further investigation.

Furthermore, some simple modifications will allow us to examine other physically interesting scenarios.
In particular, modifying the asymptotic behaviour of the black hole metric to be either de Sitter or anti-de Sitter space will allow us to study the more realistic evolution of a black hole embedded within our own Universe or to investigate nonequilibrium dynamics in the AdS/CFT correspondence, respectively.
Given the novel new dynamics that occurs already in the simple setup considered here,
there may be exciting new dynamics to understand within these other arenas.

\ack{JB and HVP are grateful for the hospitality of the Centro de Ciencias de Benasque Pedro Pascual, where the final stages of this work were completed.}
\funding{JB and HVP are supported by the European Research Council under the European Community’s Seventh Framework Programme (FP7/2007-2013) / ERC grant agreement no 306478-CosmicDawn. This work was partially enabled by funding from the UCL Cosmoparticle Initiative.}
\competing{We have no competing interests.}
\aucontribute{DT derived the equation of motion for the system and developed the numerical framework for solving it, produced the results provided in this article, contributed to the analysis and interpretation of the results and drafted the article.
JB contributed to the numerical implementation.
JB and HVP contributed to conception and design of this work, the interpretation and analysis of the results and the revision and editing of the article.
All authors read and approved the manuscript.}

\bibliography{accretion}
\bibliographystyle{vancouver}

\begin{appendices}
\renewcommand{\thesection}{\Alph{section}}
\numberwithin{equation}{section}

\section{Equation of motion}\label{app:EOM}
The action for a scalar field minimally coupled to gravity is given by
\begin{equation}
    S_\phi=\int d^4x\sqrt{-g}\left(-\frac{1}{2}g^{\mu\nu}\nabla_\mu\phi\nabla_\nu\phi-V(\phi)\right)\,,
    \label{equ:action_a}
\end{equation}
where $\phi$ is the scalar field, $V(\phi)$ is the scalar field potential, $g^{\mu\nu}$ is the background metric and $g\equiv {\rm det}\left(g^{\mu\nu}\right)$ is its determinant. 
Varying this action with respect to the scalar leads to the evolution equation for a scalar field $\phi$ coupled to a gravitational field with a metric $g_{\mu\nu}$ (with signature $(-,+,+,+)$):
\begin{equation}
  \Box\phi - \frac{\p V}{\p\phi} = \frac{1}{\sqrt{-g}}\p_\mu\left(\sqrt{-g}g^{\mu\nu}\p_\nu\phi \right) - \frac{\p V}{\p\phi} = 0\,.
  \label{equ:EOM_a}
\end{equation}
Substituting the Schwarzschild metric
\begin{equation}
  ds^2 = \left(1-\frac{2\alpha}{r}\right)dt^2 + \left(1-\frac{2\alpha}{r}\right)^{-1}dr^2 + r^2d\Omega^2
  \label{equ:schw_sln_a}
\end{equation}
into this equation, we obtain the equation of motion of a scalar field in a static spherically symmetric background
\begin{subequations}
\begin{align}
    \dot{\phi}&=\pi \, ,\\
    \dot{\pi}&= -\frac{g^{rr}}{g^{tt}}\frac{\p^2\phi}{\p r^2}-\left(\frac{2}{r}\frac{g^{rr}}{g^{tt}}+\frac{1}{g^{tt}}\p r(g^{rr})\right)\frac{\p\phi}{\p r}+\frac{1}{g^{tt}}\frac{dV}{d\phi}\,,
    \label{equ:eom_scalar_a}
\end{align}
\end{subequations}
with the relevant metric terms
\begin{subequations}
\begin{align}
    g^{tt}&=-\left(1-\frac{2\alpha}{r}\right)^{-1}\,,\qquad\qquad \sqrt{-g}=r^2\sin\theta\,,\\
    g^{rr}&=\left(1-\frac{2\alpha}{r}\right)\,, \qquad\qquad\qquad \p_rg^{rr}=\frac{2\alpha}{r^2}\,.
    \label{equ:met_elem_a}
\end{align}
\end{subequations}

As explained in the main text, it is numerically more convenient to work in the radial tortoise coordinate $\rs$, defined such that
\begin{equation}
\frac{d\rs}{dr}=\left(1-\frac{2\alpha}{r}\right)^{-1}\,,
\end{equation}
which maps the interval $r\in(2\alpha,\infty)$ to $\rs\in(-\infty,\infty)$. 
Explicitly, we have
\begin{equation}
    \rs=r+2\alpha\log\left(\left|\frac{r}{2\alpha}-1\right|\right)\,.
\end{equation}
The metric can be transformed to these coordinates
\begin{equation}
    ds^2=-\left(1-\frac{2\alpha}{r}\right)dt^2+\left(1-\frac{2\alpha}{r}\right)d{\rs}^2+r^2d\Omega^2 \, ,
\end{equation}
which leads to the following equations of motion
\begin{subequations}
\begin{align}
 \dot{\phi} &= \pi\,,\\
 \dot{\pi} &= T_1(r)\frac{\p^2\phi}{\p{\rs}^2} + T_2(r)\frac{\p\phi}{\p\rs} + T_3(r)V'(\phi)\,.
\end{align}
\end{subequations}
The $r$-dependent terms $T1$, $T2$ and $T3$ are
\begin{subequations}
\begin{align}
T_1(r)&=1\,,\\
T_2(r)&=\frac{2}{r}\left(1-\frac{2\alpha}{r}\right)\,,\, \mathrm{and} \\
T_3(r)&=-\left(1-\frac{2\alpha}{r}\right)\,.
\end{align}
\end{subequations}

\section{Numerical implementation}\label{app:num}
\subsection{Dimensionless units}
In our numerical code, we work in dimensionless variables (denoted by an overbar $\bar{{}}$):
\begin{align}
  \bar{\phi} &= \frac{\phi}{\Lambda}\,, \\
  \bar{x} &= \xScl x \,, \\
  \bar{t} &= \tScl t \, .
\end{align}
Setting $c = 1$ selects $\xScl = \tScl$. 
We now work directly with the equations of motion associated with the metric,
\begin{equation}
      ds^2  = -\left(1 - \frac{2G\mBH}{r} \right)dt^2 + \left(1 - \frac{2G\mBH}{r} \right)d\rTort^2 + r^2d\Omega^2\,.
\end{equation}
Here, the tortoise coordinate is related to the areal coordinate of the Schwarzschild metric via
\begin{equation}
  \rs = 2G\mBH\ln\left(\left|\frac{r}{2G\mBH} -1 \right|\right) + r \,.
\end{equation}
The equation of motion for the scalar field is
\begin{equation}
  \frac{\partial^2\phi}{\partial t^2} - \frac{\partial^2\phi}{\partial\rTort^2} -\frac{2}{r}\left(1-\frac{2 G\mBH}{r}\right)\frac{\partial\phi}{\partial\rTort} + \left(1-\frac{2 G\mBH}{r} \right)\frac{\partial V}{\partial\phi} = 0\,.
\end{equation}
In terms of dimensionless variables, we have
\begin{equation}
  \label{eqn:dim-eom}
  \frac{\partial^2\bar{\phi}}{\partial\bar{t}^2} - \frac{\partial^2\bar{\phi}}{\partial\bar{\rTort}^2} - \frac{2}{\bar{r}}\left(1-\frac{2\tScl G\mBH}{\bar{r}}\right)\frac{\partial\bar{\phi}}{\partial\bar{\rTort}} + \left(1-\frac{2\tScl G\mBH}{\bar{r}} \right)\frac{\partial\bar{V}}{\partial\bar{\phi}} = 0\,,
\end{equation}
where we have defined
\begin{equation}
  \bar{V} = \frac{1}{\tScl^2\fScl^2}V \, .
\end{equation}
For a general polynomial potential, we have
\begin{equation}
  V(\phi) = \sum_i \lambda_iM_{S}^{4-i}\phi^i = \sum_i \lambda_i M_{S}^4\left(\frac{\fScl}{M_{S}}\right)^i\bar{\phi}^i\,,
\end{equation}
with the $\lambda_i$ denoting dimensionless coefficients.
The dimensionless potential for numerical simulations is then
\begin{equation}
  \bar{V}(\bar{\phi}) = \sum_i \lambda_i \frac{M_S^4}{\tScl^2\fScl^2} \left(\frac{\fScl}{M_S}\right)^i\bar{\phi}^i\,.
\end{equation}
In order to fix the form of the differential part of the equation of motion, it is convenient to choose
\begin{equation}
  \label{eqn:mass-scale}
  \tScl = \frac{1}{G\mBH} = \mpl\left(\frac{\mpl}{\mBH}\right)\,,
\end{equation}
where $\mpl = G^{-1/2}$ is the Planck mass.
Our dimensionless equation of motion~\eqref{eqn:dim-eom} now has the form
\begin{equation}\label{eqn:eom-final}
  \frac{\partial^2\bar{\phi}}{\partial\bar{t}^2} - \frac{\partial^2\bar{\phi}}{\partial\bar{\rTort}^2} - \frac{2}{\bar{r}}\left(1-\frac{2}{\bar{r}}\right)\frac{\partial\bar{\phi}}{\partial\bar{\rTort}} + \left(1-\frac{2}{\bar{r}} \right)\frac{\partial\bar{V}}{\partial\bar{\phi}} = 0\,,
\end{equation}
so that all of our model dependence and parameter dependence is encoded in the form of $\bar{V}$.

In order to solve~\eqref{eqn:eom-final} numerically, we must first decide on a method to discretise the spatial coordinate $\rs$ to compute spatial derivatives, and the time coordinate $t$ to provide a time-stepping procedure.

For the spatial discretisation, we use a pseudospectral approximation.
First, we expand the field (as a function of the tortoise coordinate $\rs$) in a series of rational Chebyshev functions of the doubly infinite interval ($TB_i(r)$)
\begin{equation}\label{eqn:basis-expansion}
  \phi(\rs,t) = \sum_{i=0}^{O} c_i(t)TB_i(\rs) = \sum_{i=0}^{O} c_i(t)B_i\left(\frac{\rs}{\sqrt{L^2+\rs^2}}\right) \, ,
\end{equation}
where $O$ is the order of the expansion, $B_i$ are the Chebyshev polynomials, and $L$ is an adjustable parameter.
Through this work we have set $\mu L = 10$, which we found was sufficient for the field evolutions studied here.
We next choose a set of $O+1$ gridpoints at which to store the values of $\phi$, which are given by the zeros of $TB_{O+1}$.
The coefficients $c_i$ can then easily be obtained through numerical quadrature.
Given the  $c_i$'s, derivatives of $\phi$ are then computed by direct differentiation of the basis functions $TB_i$.
Provided that the order of the expansion~\eqref{eqn:basis-expansion} is high enough to properly resolve all of the spatial structure in $\phi$, this leads to a rapidly (often exponentially) convergent approximation for the field $\phi$.

As for the time evolution, we use a tenth-order accurate Gauss-Legendre method,
which is based on repeated application of a low-order pseudospectral approximation for the time-direction.
Details on these methods may be found in Ref.~\cite{Butcher1964} or Appendix B of Ref.~\cite{Braden:2014cra}
In addition to its high accuracy, this method is also symplectic, time-reversible, and A-stable, thus providing a set of desirable numerical properties.

\subsection{Inverting tortoise coordinates} 
The equation that we are solving \eqref{equ:eom_tor} contains $r$\textendash dependent terms, but our equations are being solved in the tortoise coordinates $\rs$.
The relation between the tortoise, $\rs$, and areal, $r$, coordinates was defined as
\begin{equation}
    \rs=r+2\log\left(\left|\frac{r}{2}-1\right|\right),
\end{equation}
which needs to be inverted to evaluate each $r\in(2\alpha,\infty)$ corresponding to a point $\rs\in(-\infty,\infty)$.
This can be done with a root-finding algorithm, for which an initial guess for the relation had to be supplied.
We started by setting the value of $r$ on the first grid point, $r(1)=2\alpha$ and then proceeded, from $r(2)$ to $r(N)$, to estimate the values at each step as functions of $\rs$ at that step and $r$ at the previous step as
\begin{equation}
r(j)=
\begin{cases}
    2\left[1+\exp\left(\frac{\rs(j)-r(j-1)}{2}\right)\right]\qquad\,\,\,\, \text{if}\,\, r(j-1) > 1.2\times\rs(j); \\
    \rs(j) - 2\log\left(\left|\frac{r(j-1)}{2}-1\right|\right)\qquad\qquad\text{otherwise},
\end{cases}
\end{equation}\\
where $j=2,3,...,N$.
We have estimated this as a piecewise function, because as $r$ increases $\rs$ approaches $r$ and the estimation using the exponential of their difference will no longer give accurate results. Therefore we have set a condition, which ensures that when the values of the two coordinates get too close the estimated value of $r$ is given by the lower relation.\\
\begin{figure}[H]
    \centering
    \begin{subfigure}[t]{0.48\textwidth}
        \includegraphics[width=\textwidth]{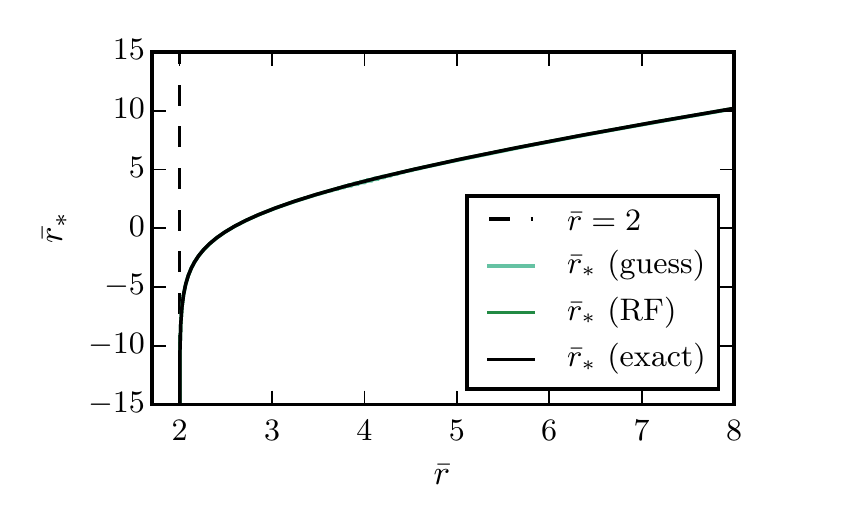}
    \end{subfigure}
    \,
    \begin{subfigure}[t]{0.48\textwidth}
        \includegraphics[width=\textwidth]{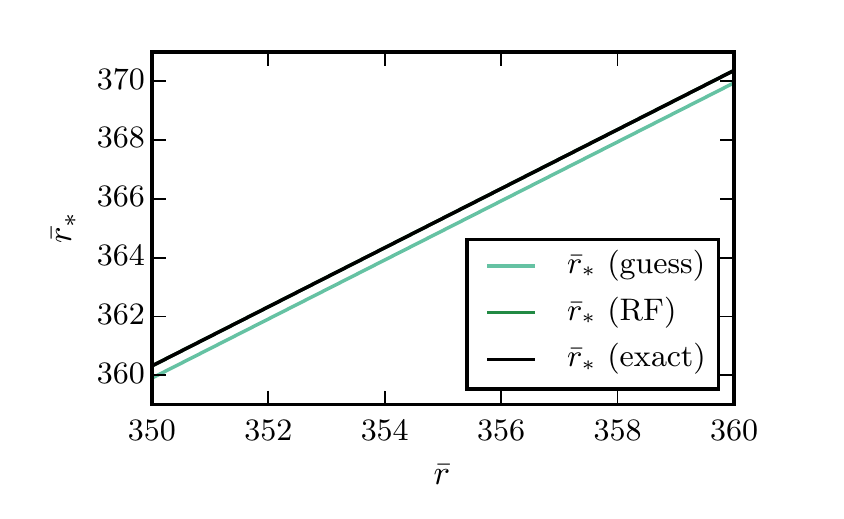}
    \end{subfigure}
    \caption{The tortoise coordinate, $\rs$, as a function of $r$ given by the exact relation (labelled as exact), our initial guess (labelled as guess) and our improved coordinate transformation after applying our root finding algorithm (labelled as RF). The dashed line marks the location of the Schwarzschild radius $\mu\,r = 2$.}
    \label{fig:r-rs}
\end{figure}
In~\figref{fig:r-rs} we show the exact function $\rs$ as a function of $r$, our initial guess described above, and our final numerical approximation using numerical root finding.
We can see that our initial guess is a good approximation to the true relationship, with a small systematic offset appearing at large $r$.  Meanwhile, our numerically obtained relationship provides an excellent approximation throughout the domain.
\end{appendices}
\end{document}